\documentclass[12pt]{iopart}
\usepackage{graphicx}

\usepackage{iopams}
\begin{document}

\title[Graphene NEMS resonator]{Probing thermal expansion of
graphene and modal dispersion at low-temperature using graphene NEMS
resonators}

\author{Vibhor Singh$^1$, Shamashis Sengupta$^1$, Hari S. Solanki$^1$,
Rohan Dhall$^1$, Adrien Allain$^1$, Sajal Dhara$^1$, Prita Pant$^2$
and Mandar M. Deshmukh$^1$}

\address{$^1$Department of Condensed Matter Physics, TIFR, Homi Bhabha
Road, Mumbai 400005 India}

\address{$^2$Department of Metallurgical Engineering and Materials
Science, IIT Bombay Powai, Mumbai : 400076, India}

\ead{deshmukh@tifr.res.in}

\begin{abstract}
We use suspended graphene electromechanical resonators to study the
variation of resonant frequency as a function of temperature.
Measuring the change in frequency resulting from a change in
tension, from 300~K to 30~K, allows us to extract information about
the thermal expansion of monolayer graphene as a function of
temperature, which is critical for strain engineering applications.
We find that thermal expansion of graphene is negative for all
temperatures between 300K and 30K. We also study the dispersion, the
variation of resonant frequency with DC gate voltage, of the
electromechanical modes and find considerable tunability of resonant
frequency, desirable for applications like mass sensing and RF
signal processing at room temperature. With lowering of temperature,
we find that the positively dispersing electromechanical modes
evolve to negatively dispersing ones. We quantitatively explain this
crossover and discuss optimal electromechanical properties that are
desirable for temperature compensated sensors.
\end{abstract}


\maketitle

\section{Introduction}
Electronic properties of graphene have been studied extensively
\cite{castronetoRMP,geim-review-future} since the first experiments
probing quantum Hall effect
\cite{kimgraphenefirst,geimgraphenefirst}. In addition to the
electronic properties, the remarkable mechanical properties of
graphene include a high in-plane Young's modulus of $\sim$1~TPa
probed using nanoindentation of suspended
graphene\cite{jimhone-graphene-nanoindent}, force extension
measurements \cite{mechanicalproperty}, and electromechanical
resonators \cite{bunchgrapheneresonator,jimhonegraphene-resonator,
bachtoldsuspendedgrapheneimaging}. NEMS (nano electromechanical
systems) devices using nanostructures like carbon nanotubes
\cite{veranature,AKHuttelnanoletters,sciencevanderzant,sciencebachtold,Mass1b,Mass1c,Mass4+Mixer1},
nanowires \cite{yang-roukes} \cite{hari-resonator} and bulk
micromachined structures
\cite{monolithicsculpted,Kottahausnewnatphys,akshaynaik} offer
promise of new applications and allow us to probe fundamental
properties at the nanoscale. NEMS \cite{ekinci_review} based devices
are ideal platforms to harness the unique mechanical properties of
graphene. Electromechanical measurements with graphene resonators
\cite{bunchgrapheneresonator,jimhonegraphene-resonator} suggest that
with improvement of quality factor ($Q$), graphene based NEMS
devices have the potential to be very sensitive detectors of mass
and charge. Additionally, the sensitivity of graphene to chemical
specific processes \cite{geim-gas-sensor} offers the possibility of
integrated mass and chemical detection. The large surface-to-mass
ratio of graphene offers a distinct advantage over other
nanostructures for such applications. In order to better understand
the potential of graphene based electrically actuated and detected
resonators \cite{jimhonegraphene-resonator} and the challenges in
realizing strain-engineered graphene devices
\cite{castro-neto-strainengineering,geim-graphene-strained}, we
experimentally measure the coefficient of thermal expansion of
graphene ($\alpha_{graphene}(T)$) as a function of temperature. Our
measurements indicate $\alpha_{graphene}(T)$~$<0$ for
30~K~$<T<$~300K and larger in magnitude than theoretical prediction
\cite{expansiontheory}. We also probe the dispersion, or the
tunability, of mechanical modes using the DC gate voltage at low
temperatures and find that the thermal expansion of graphene,
built-in tension and added mass play an important role in changing
the extent of tunability of the resonators \cite{nguyen-tunable} and
the resonant frequency -- parameters that are critical for various
applications. Additionally, measuring temperature dependent
mechanical properties \cite{lau-graphene-rippled} of suspended
structures down to low temperatures will give insight into strain
engineering of graphene based devices
\cite{castro-neto-strainengineering,geim-graphene-strained} and also
help in understanding the role of rippling in degrading carrier
mobility at low temperatures. Our experiments probe these issues in
detail and suggest that monitoring the resonant frequency as a
function of temperature could provide important information about
nanoscale stress that is useful for probing phase transition.

\section{Experimental details}

\subsection{Device fabrication}
To fabricate monolayer graphene electromechanical resonators, we
suspended graphene devices using the previously reported
\cite{evaandrei-suspended,bolotinsuspended-first,bolotinsuspended-prl,
jimhonegraphene-resonator} process which starts with micromechanical
exfoliation of graphene from graphite crystal
using an adhesive tape on a 300 $\mu$m
thick degenerately doped silicon wafer coated with 300~nm thick
thermally grown SiO$_2$. Following the location of monolayer
graphene flakes using optical microscopy, electron beam lithography
is used to pattern resist for fabricating electrodes for electrical
contact. The electrodes are fabricated by evaporation of 10~nm of
chromium and 60~nm of gold following patterning of resist. To
release the graphene from the SiO$_2$ substrate, a dilute
buffered-HF solution is used to selectively etch an area around the
graphene device by masking the rest of the substrate using polymer
resist. Following an etch for 5 min 30~sec that results in a 170 nm
deep trench in SiO$_2$, the device is rinsed in DI water and
isopropanol. Critical point drying, to prevent collapse of the
device due to surface tension, is the final step in fabrication of
suspended graphene devices. A scanning electron microscope (SEM)
image of a suspended graphene device is shown in Figure 1a.

\subsection{Measurement scheme}
\begin{figure}
\begin{center}
\includegraphics[width=65mm]{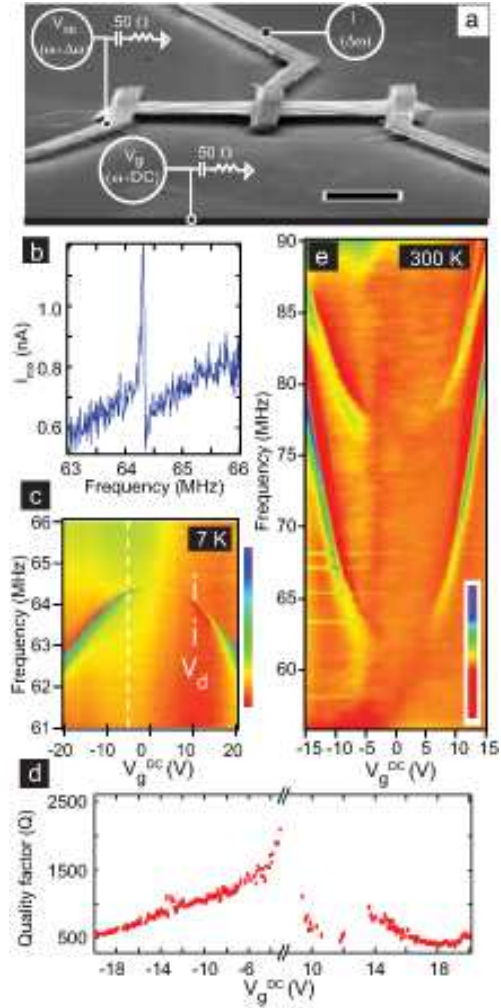}
\end{center} \caption{\label{fig:figure1} a) Tilted angle scanning
electron microscope (SEM) image of a suspended monolayer graphene
device and the electrical circuit for actuation and detection of the
mechanical motion of the graphene membrane. The scale bar indicates
a length of 2$\mu$m. b) A plot of the mixing current I$_{mix}(\Delta
\omega)$ as a function of frequency $f(=\frac{\omega}{2 \pi})$ at
7~K with the DC gate voltage $V_g^{DC}= -5V$ for device (D1). The
sharp feature in the mixing current corresponds to the mechanical
resonance.  c) Colorscale plot of the mixing current as function of
frequency $ f $ and DC gate voltage $V_g^{DC}$ at 7~K for device
(D1). The dashed line shows the position of the line scan shown in
Figure \ref{fig:figure1}b. The dot-dash line indicates the position
of the Dirac peak for the graphene device. The measured dispersion
of this mode of the device is negative. The maxima (blue) and minima
(red) of the colorbar correspond to 2.6nA and 0nA respectively. d)
Plot of the measured Q for the data shown in Figure
\ref{fig:figure1}c. The Q drops as $|V_g^{DC}|$ increases and around
the Dirac point the Q shows a dip. e) Colorscale plot of
I$_{mix}(\Delta \omega)$ as a function of frequency
$f=\frac{\omega}{2 \pi}$ and the DC gate voltage. $V_g^{DC}= 13$V
for device (D2) at 300K. Two positively dispersing modes of the
device are seen.} \end{figure}

The electrical actuation and detection is done by using the
suspended graphene device as heterodyne-mixer
\cite{jimhonegraphene-resonator,veranature,bachtoldmasssensing}. The
scheme for electrical actuation and detection is also shown in
Figure 1a superimposed on the SEM image of the device. We use
electrostatic interaction between the graphene membrane and the
back-gate electrode to actuate the motion in a plane perpendicular
to the substrate. A radio frequency (RF) signal $V_g(\omega)$ and a
DC voltage $V_g^{DC}$ are applied at the gate terminal using a
bias-tee. Another RF signal $V_{SD}(\omega +\Delta\omega)$  is
applied to the source electrode Figure \ref{fig:figure1}a. RF signal
applied at the gate $V_g(\omega)$ modulates the gap between graphene
and substrate at frequency $\omega$, and $V_g^{DC}$ alters the
overall tension and carrier density in the membrane. The amplitude
of the current through the graphene membrane at the difference
frequency ($\Delta\omega$), also called the mixing current
$I_{mix}(\Delta \omega)$, can be written as
\cite{jimhonegraphene-resonator,veranature,Mass4+Mixer1,Mixer2a,Mixer2b}
\begin{equation}
I_{mix}(\Delta \omega)=\frac{1}{2}\frac{dG}{dq}(\frac{dC_g}{dz}
z(\omega) V_g^{DC} + C_g V_g(\omega))V_{SD}(\omega + \Delta \omega),
\label{eq:equation1}
\end{equation}
where $G$ is the conductance of the graphene device, $q$ is the
charge induced by the gate voltage, $C_g$ is the capacitance between
the gate electrode and graphene, $z(\omega)$ is the amplitude of
oscillation at the driving frequency $\omega$ along the $z$-axis
(perpendicular to the substrate). The difference frequency signal
(at $\Delta \omega$) arises from the product of the modulation
signals of $V_{SD}(\omega + \Delta \omega)$ and $G(\omega)$. At the
mechanical resonance of the membrane, the first term in
Equation~\ref{eq:equation1} contributes significantly and the second
term which does not depend on the mechanical motion of the graphene
membrane provides a smooth background.

\subsection{Results and discussions}
Figure \ref{fig:figure1}b shows the result of such a measurement at
7K for a suspended graphene device (D1) while $V_g^{DC}$ is set at
-5V. Using a Lorentzian lineshape for the resonance curve,
\cite{veranature,Vera_thesis} we can extract the $Q \sim 1500$ of
the resonator (variation of $Q$ with temperature is shown in
supplementary information). Figure \ref{fig:figure1}c shows the
colorscale plot of $I_{mix}(\Delta \omega)$ as a function of
$V_g^{DC}$ and $f=\omega/2 \pi$ at 7~K. The resonant frequency is
lowered as the magnitude of $V_g^{DC}$ is increased - mechanical
mode disperses negatively with $|V_g^{DC}|$. This is well understood
in terms of mode-softening due to the capacitive contribution to the
energy of resonator \cite{kozinsky,hari-resonator}. The Dirac peak
for our device is shifted ($V_D = 13V$) from the expected position
(0~V) due to unintentional doping during the fabrication process
\footnote[1]{A shift in the Dirac peak away from $V_g^{DC}=0V$ is
desirable in our experiments because the actuation efficiency is
feeble at $V_g^{DC}=0V$ and that would make the observation of
physics near the Dirac point inaccessible for electromechanical
measurements.}. Also seen in Figure \ref{fig:figure1}c is the
information regarding the amplitude of the mixing current. The
overall amplitude of the mixing current at resonance scales with
$V_g^{DC}$ (first term in Equation \ref{eq:equation1}) and as a
result the amplitude of the mixing current is very small near
$V_g^{DC}=0V$. However, at $V_g^{DC}=13$V, the amplitude of the
mixing changes abruptly due to the Dirac peak ($V_D$)
\cite{kimgraphenefirst,geimgraphenefirst}. At this point, the term
$\frac{dG}{dq}$ appearing in Equation \ref{eq:equation1} becomes
zero. Figure \ref{fig:figure1}d shows the quality factor ($Q$)
dependance on $V_g^{DC}$ calculated from the data shown in Figure
\ref{fig:figure1}c. The quality factor of the device is largest
around $V_g^{DC} = 0V$ and decreases as the gate voltage is swept
away from zero. Also, at low temperatures, $f_0$ decreases with the
increase in $|V_g^{DC}|$ and $Q$ follows a similar trend. As
negative dispersion -- caused by softening of spring constant -- is
accompanied by the movement of graphene towards the gate electrode,
it leads to an increase in the modulated capacitance and results in
larger dissipative current \cite{kozinsky}. In the neighborhood of
$V_g^{DC} = V_D$,  $Q$ is smaller than the value expected from the
trend from higher $V_g^{DC}$ side. One possible reason for this
behavior might be that in the vicinity of the Dirac point, large
device resistance leads to increased ohmic losses
\cite{Vera_thesis}. An additional mechanism that can result in
dissipation is the mechanical deformation modifying the distribution
of charge puddles of electrons and holes in the graphene sheet.
These phenomena warrant further detailed investigation
\cite{yacoby-puddles,crommie-puddles}. Such a measurement could be
used to study charge inhomogeneity. Figure \ref{fig:figure1}e shows
similar measurement for a device (D2) at 300K. Here, we observe two
mechanical modes which disperse positively because of the increase
in tension in the graphene membrane due to the electrostatic
attraction from the back-gate. Most of our graphene devices exhibit
an in-built tensile stress at 300K due to the fabrication process.
Multiple ripples on the membrane are also seen in Figure 1a
\cite{lau-graphene-rippled}. To better understand the modal
dispersion quantitatively, we model the graphene membranes within
the elastic continuum regime and in a limit where tension in the
membrane dominates over the flexural rigidity
\cite{jimhone-graphene-nanoindent,bunchgrapheneresonator,graphene-resonator-continuum}.
As a result, the resonant frequency $f_0$ of the graphene membrane
can be written as

\begin{equation}
f_0(V_g^{DC})=
\frac{1}{2L}\sqrt{\frac{\Gamma(\Gamma_0(T),V_g^{DC})}{\rho t w}},
\label{eq:equation2}
\end{equation}
where $L$ is the length of the membrane, $w$ is the width, $t$ is
the thickness, $\rho$ is the mass density, $\Gamma_0(T)$ is the
in-built tension and $\Gamma$ is the tension at a given temperature
$T$ and $V_g^{DC}$. The functional form of
$\Gamma(\Gamma_0(T),V_g^{DC})$ is dependent on the details of the
model used to take into account the electrostatic and elastic
energies. (Our model is described in detail in the supplementary
information.) Resonant frequency at zero gate voltage is given by
$f_0(0)= \frac{1}{2L}\sqrt{\frac{\Gamma_0(T)}{\rho t w}}$;
therefore, from $f_0(0)$ an independent estimation of $\Gamma_0(T)$
and $\rho$ is not possible. However, resonant frequency dispersion
with gate voltage $f_0(V_g^{DC})$ allows us to estimate $\rho$ and
$\Gamma_0(T)$ simultaneously. Using fits based on the continuum
model (calculations described in supplementary information) we
estimated the mass density $\rho= 7.4 \rho_{graphene}$ and the
in-built tension $\Gamma_0(300K)$ to be $68.6 $~nN for the
fundamental mode, where $\rho_{graphene}$ is the mass density of
pristine graphene for the device data shown in Figure
\ref{fig:figure1}e. We attribute this extra mass and in-built
tension to the resist residue that can get deposited on the graphene
membrane during the fabrication process and these values of mass
density are similar to ones reported by Chen \emph{et al.}
\cite{jimhonegraphene-resonator}. We note that the dispersion of the
two modes shown in the Figure \ref{fig:figure1}e is different and
our modal calculations give $\rho= 7.3 \rho_{graphene}$ and
$\Gamma_0(300K)= 107.6$~nN for the upper mode. However, the simple
model of a rectangular membrane predicts higher order modes at much
higher frequencies -- this is a limitation of the simple assumptions
we have made. The presence of other resonant modes at small
intervals of frequency $f$ seen in Figure \ref{fig:figure1}e (also
seen in Figure \ref{fig:figure4}a) indicates a deviation from this
simplified picture of membrane under tension due to either
edge-modes \cite{bachtoldsuspendedgrapheneimaging}, or due to the
rippling/curling of the membrane
\cite{lau-graphene-rippled,cornellcurledgraphene}. In the later part
of this report, we discuss in detail the reasons for positive or
negatively dispersing modes and temperature induced change in
dispersion.

\section{Probing thermal expansion of graphene}
\begin{figure}
\begin{center}
\includegraphics[width=85mm]{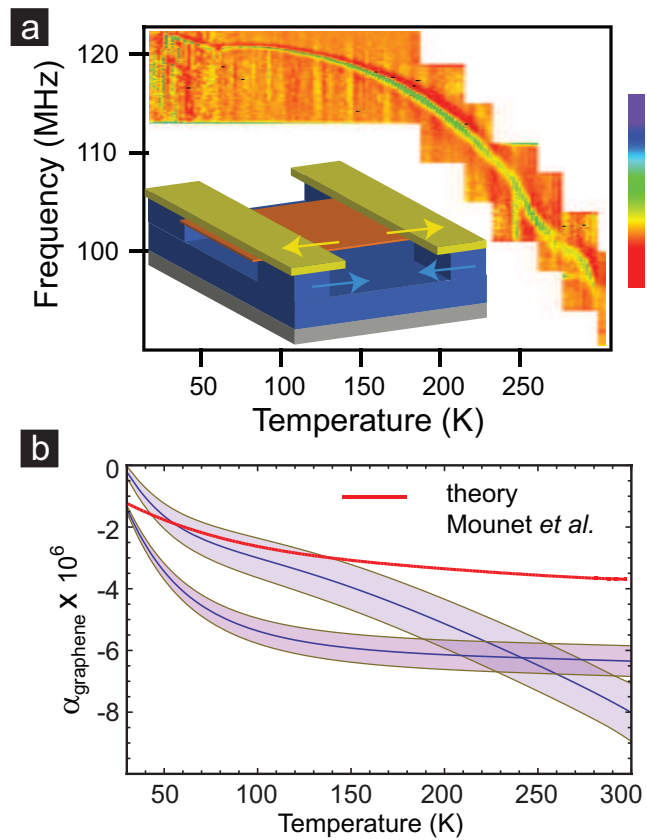}
\end{center}
\caption{\label{fig:figure2} a) Plot showing the evolution of the
resonant frequency of a mode for device (D2) as a function of
temperature for $V_g^{DC}=15$~V. Inset shows the schematic of all
the strains external to the suspended graphene membrane as the
device is cooled below 300~K. b) The plot of expansion coefficient
of graphene as a function of temperature. Data from two different
devices together with theoretical prediction of N. Mounet \emph{et
al.} \cite{expansiontheory}. The shaded area represents the errors
estimated from the uncertainty of the length of the flake, width of
the electrode and Young's modulus of graphene. }
\end{figure}

We now consider how the resonant frequency ($f_0$) evolves as a
function of the temperature. Figure \ref{fig:figure2}a shows the
result of an evolution of a mode as a function of temperature at
$V_g^{DC}=15$~V \footnote[2]{Data acquisition was done during a
single sweep over twelve hours to allow the resonator to equilibrate
and the window of acquisition window automatically adjusted to
follow the resonance.}. The resonant frequency increases as the
device is cooled from room temperature. This increase has been seen
in all the devices we have studied.
The degree of frequency shift varies from one device to another
depending on the device geometry. The origin of this frequency shift
with temperature is the increase in tension in graphene due to the
expansion/contraction of substrate, gold electrodes and graphene.
The frequency shift can be understood by taking into account the
contribution of various strains as the device is cooled below 300~K.
The three main contributions are -- firstly, the thermal strain in
unconstrained graphene $\epsilon_{graphene}(T) = \int_T^{300}
\alpha_{graphene}(t) dt$ \footnote[3]{All through this report we
measure the strain relative to the strain at room temperature
(300~K).} due to the coefficient of thermal expansion of graphene
$\alpha_{graphene}(T)$, second, the thermal strain due to the gold
electrodes $\epsilon_{gold}(T) = \int_T^{300} \alpha_{gold}(t) dt$
and lastly the contribution of the strain induced by the substrate
$\epsilon_{substrate}(T)$; here $\alpha_{gold}(T)$ is the
coefficient of thermal expansion for gold
\cite{goldthermalexpansion}. The strain in gold electrodes plays an
important role due to the geometry of the device. The under-etch
that releases the graphene membrane also etches under the graphene
covered by the electrodes -- resulting in the graphene membrane
being suspended off the gold electrodes
\cite{jimhonegraphene-resonator} (see supplementary information for
SEM image). The geometry of the resulting device is shown in the
inset to Figure \ref{fig:figure2}a. The elastic strain in the curved
substrate can be calculated using Stoney's equation
\cite{freundsuresh}. Its contribution to the net strain in the
graphene membrane is very small (see supplementary information) and
therefore upon cooling, change of tension in the graphene is due to
contraction of gold electrode and expansion/contraction of graphene.
We assume that Young's modulus of graphene does not vary
significantly over the temperature range
\cite{Young'sModulusTemperatureDepandance}. At the interface of gold
electrodes supporting the graphene membrane, the net force must
balance to zero; however, the stresses are different considering the
crossectional area of gold electrodes ($\sim$500nm $\times $ 60nm)
and graphene ($\sim$500nm $ \times $ 0.3nm). The large difference in
the cross-sectional area implies that the effective stiffness of
gold electrodes is large compared to the stiffness of graphene. As a
result, to a very good approximation, the total elastic strain at a
given temperature in graphene that is confined by ``rigid" gold
electrodes is $\epsilon_{graphene clamped}=\epsilon_{graphene}(T)
 +\epsilon_{substrate}(T)
  - \epsilon_{gold}(T)
\frac{w_{electrode}}{L}$, where $w_{electrode}$ is the average of
the width of gold electrodes holding the suspended flake. The change
in tension in the membrane can be written as a function of
temperature as $\Delta \Gamma_0(T) = wt\epsilon_{graphene
clamped}(T) E_{graphene}$, where $E_{graphene}$ is Young's modulus
of graphene. Measuring the tension $\Gamma_0(T)$ as a function of
temperature offers a way to track the thermal strain in graphene
membrane. Figure \ref{fig:figure2}a shows the evolution of resonant
frequency as a function of temperature from a device (D2) at
$V_g^{DC}=15$~V where the increase in frequency is largely due to
the contraction of the gold electrodes. However, this rate of
increase of resonant frequency is significantly reduced due to the
negative $\alpha_{graphene}$ for all T$<300$~K from the case of
frequency change including only gold's contraction. Using such a
measurement of frequency shift while assuming uniform expansion of
all the materials and using Equation~\ref{eq:equation2}, it is
possible to calculate the expansion coefficient from the frequency
at $V_g^{DC}=0$~V as
\begin{equation}
\fl \alpha_{graphene}(T)= - 2 f_0(0) \frac{df_0(0)}{dT}
\times\frac{(2 L) ^2 \rho}{E_{graphene}} + \frac{d}{dT} (
\epsilon_{substrate}(T) - \epsilon_{gold}(T) \frac{w_{electrode}}{L}
) \label{eq:equation4}
\end{equation}
Figure \ref{fig:figure2}b shows the result of calculating
$\alpha_{graphene}$ for two devices using this analysis and
comparison with the theoretical calculation for $\alpha_{graphene}$
by N. Mounet \emph{et al.} \cite{expansiontheory} . We find that
$\alpha_{graphene}$ is negative and its magnitude decreases with
temperature for T$<300$~K. At room temperature, $\alpha_{graphene}
\sim  -7 \times 10^{-6} K^{-1}$ , which is similar to the previously
reported values measured by others
\cite{jimhonegraphene-resonator,lau-graphene-rippled}. At 30~K,
$\alpha_{graphene} \sim  -1 \times 10^{-6} K^{-1}$. The deviation of
$\alpha_{graphene}$ from the theoretically predicted values can
possibly be due to the presence of the impurities on graphene
membrane. The knowledge of $\alpha_{graphene}$ is essential for the
fabrication of the devices intended for strain engineering
applications
\cite{castro-neto-strainengineering,geim-graphene-strained}.  Strain
engineering of graphene devices at low-temperatures using this
picture can improve device performance, for example by enabling
temperature compensation \cite{temperature-compensated}. A simple
design rule for width of electrodes to achieve temperature
compensation in the vicinity of a temperature $T_0$ is to ensure
that the
$w_{electrode}=L\times|\frac{\alpha_{graphene}(T_0)}{\alpha_{gold}(T_0)}|$
(using equation \ref{eq:equation4}). Additionally, these
measurements indicate that using NEMS resonators for measuring
internal stress of nanowires of metal and phase change materials as
a function of temperature, using phase locked loop (PLL) technique
\cite{akshaynaik}, can provide useful information about stresses and
strains in individual nanostructures.

Here, it is important to point out that, residue on the graphene
sheet can also affect the calculations of the coefficient of thermal
expansion. This can possibly be one reason for its deviation from
the theoretically predicted values. In our calculation, we have
assumed that presence of residue does not affect the expansion of
graphene, which may not be the complete description of the
experimental system. However, these values of $\alpha_{graphene}$
still remain valuable for device engineering as the complete
elimination of the resist residue from the device is difficult.

%

\section{Modal dispersion of graphene NEM resonator}
\begin{figure}
\begin{center}
\includegraphics[width=85mm]{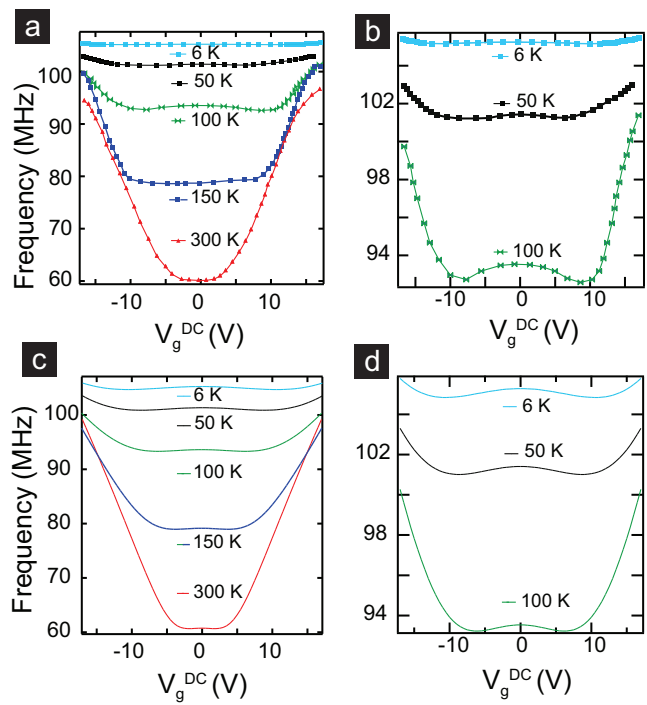}
\end{center} \caption{\label{fig:figure3} a) Measured dispersion of
an electromechanical mode as function of temperature for device
(D3). b) Closeup of the experimentally measured low-temperature data
showing the non-monotonic dispersion. c) Modeling of the dispersion,
incorporating thermal expansion of graphene, at various temperatures
by varying temperature and the parameter $\lambda$. d) Closeup of
the calculated low-temperature data showing the non-monotonic
dispersion similar to the experimentally measured dispersion. }
\end{figure}

We next consider the dispersion of the resonant modes with
temperature. Figure \ref{fig:figure3}a shows the dispersion of
resonant mode for a device(D3) at various temperatures down to 6K.
Upon cooling, the resonant frequency increases and its tunability
using gate voltage is reduced, an undesirable feature. Further, at
low temperatures, resonance frequency first decreases with
$|V_g^{DC}|$ and then it increases (Similar evolution of modal
dispersion with $V_g^{DC}$ has been observed by Chen et al.
\cite{jimhonegraphene-resonator}.). We now try to understand the
evolution of the modal dispersion as a function of temperature.
The two limiting cases of modal dispersion -- pure positively and
negatively dispersing modes can be respectively understood by
considering the limits where the tension (or in the case of flexural
modes mechanical rigidity) dominates with increasing $V_g^{DC}$ and
the case when capacitive pulling leading to the softening of the
spring constant dominates
\cite{kozinsky,hari-resonator,Vera_thesis}. A simple model
\cite{hari-resonator} to understand the intermediate regime where
these two interactions compete is to model the resonator with an
intrinsic spring constant $K_i=k+\alpha (V_g^{DC})^2 + \beta
(V_g^{DC})^4 + H.O.(V_g^{DC})$ with $\alpha$ and $\beta$ as
constants of the system (when $\beta > 0$, this is consistent with
the positive dispersion of modes with an increased $V_g^{DC}$). A
second contribution due to the electrostatic interaction (capacitive
coupling), softens the intrinsic spring constant $K_i$ so that the
effective spring constant
\begin{equation} K_{eff}=K_i -
\frac{1}{2}(V_g^{DC})^2 \frac{d^2C_g}{dz^2}. \label{eq:springSoft}
\end{equation} with $C_g$ being
the capacitive coupling of the membrane with gate electrode. As
$V_g^{DC}$ is varied the $K_{eff}$ varies and the modal dispersion
changes from a negatively dispersing mode to a positively dispersing
mode at large $V_g^{DC}$ -- the value of $V_g^{DC}$ at which the
crossover happens is a measure of the relative contribution of
capacitive and elastic energies. This phenomenological model
describes the generalized modal dispersion and can be connected to
the properties of the resonator (described in supplementary
information). $\alpha$ and $\beta$ depend on temperature through the
coefficient of thermal expansion, and also on the elastic constant
of graphene; as a result they change with temperature, leading to
the evolution of modal dispersion as a function of temperature.
Additionally the capacitive contribution to energy is expected to
change as a function of temperature. The observed modal dispersion
behavior at all temperature can be qualitatively understood from the
contraction of the suspended gold electrodes with temperature, which
leads to increased tension in the membrane. This increase in
in-built tension is accompanied by the reduction in tunability with
$V_g^{DC}$ (model described in the the supplementary information).
Results of our calculation (Figure \ref{fig:figure3}c) clearly show
that we can successfully model the temperature evolution of
dispersion, using only a single fit parameter in our model.
 It is
critical to understand the origin of observed dispersion of resonant
frequency as a function of temperature due to the desirable property
of resonator -- its tunability. Additionally, if the loss mechanisms
are frequency independent such tunability can increase the $Q$ of
the system \cite{Vera_thesis}.

\begin{figure}
\begin{center}
\includegraphics[width=85mm]{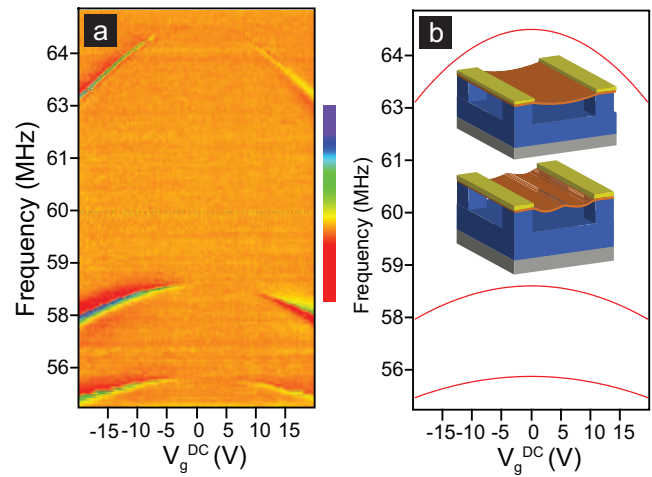}
\end{center}
\caption{ \label{fig:figure4} a) Colorscale plot of I$_{mix}(\Delta
\omega)$ showing three negatively dispersing modes with varying
slopes. b) Fitted data for varying effective mass for the three
modes. The inset shows two possible modes for a uniform rectangular
membrane under tension. The effective mass is expected to be
independent of the mode -- an observation that is not experimentally
seen within the measured frequency range.}
\end{figure}

In order to use graphene electromechanical resonators for
applications like mass and charge sensing it is important to
understand the microscopic origin of the modes and to test the
suitability of continuum models
\cite{bunchgrapheneresonator,jimhonegraphene-resonator,graphene-resonator-continuum}
beyond basic properties such as dispersion. Scanned probe
measurements to image modes \cite{bachtoldsuspendedgrapheneimaging}
indicate that the origin of  these modes is complex -- for instance
some of the modes are associated with the edges of the graphene
membrane. The reason this is critical is that the notion of
effective mass of mechanical modes
\cite{ekinci_review,karabacakthesis} is associated with the spatial
distribution of the mode \footnote[4]{The effective mass $m_{eff}$
of a mode is dependent on the mode in flexural oscillations. For the
case of string under tension the $m_{eff}$ is independent of the
mode \cite{karabacakthesis}.}. We try to understand the nature of
modes in graphene resonators by studying devices with multiple
resonances in the measurement frequency window. Figure
\ref{fig:figure4}a shows two higher order modes along with
fundamental mode for a graphene resonator at 7~K. The dispersions of
these three modes are different. Though it is difficult to determine
the exact nature of these modes using our technique, a
semi-quantitative explanation is provided here. At lower
temperatures, negative dispersion can be understood in terms of
spring constant softening as discussed earlier. Following
Equation~\ref{eq:springSoft}, the effective spring constant
$K_{eff}$ can be used to calculate the modal dispersion with gate
voltage~($V_g^{DC}$),
\begin{equation}
f^2 = f_0^2- \frac{1}{8 \pi^2 m_{eff}}(V_g^{DC})^2
\frac{d^2C_g}{dz^2} \label{eq:frqSoft},
\end{equation}
where $f_0 = \frac{1}{2 \pi} \sqrt{\frac{K_i}{m_{eff}}}$ and
$m_{eff}$ is the effective mass for a given mode
\cite{ekinci_review,karabacakthesis}.

Using Equation~\ref{eq:frqSoft}, we can describe the higher modes
also by using $f_0$ and modal mass ($m_{eff}$) as the fitting
parameters. For a sheet under tension with uniform loading, the
effective modal mass ($m_{eff}$) for different modes does not
change, unlike the case of flexural modes, and is equal to $m_{eff}
= 0.785 \rho L w t $ \cite{karabacakthesis}. However, our fitting
gives different effective masses for different modes, which is
expected as these are not the higher order harmonics of the
fundamental mode (in the model described above, the higher order
modes will be integral multiples of the fundamental mode). This
suggests that the simple picture of rectangular membrane under
tension within the continuum description does not work well to
describe the system. This could be due to four main reasons: a)
non-uniform loading of the membrane (due to resist residue)
\cite{jimhonegraphene-resonator} can modify the $m_{eff}$ for
different modes, b) due to the presence of edge dependent modes
\cite{bachtoldsuspendedgrapheneimaging}, c) due to the curvature of
membrane \cite{cornellcurledgraphene} or rippling of graphene
\cite{lau-graphene-rippled} (as seen in Figure \ref{fig:figure1}a)
the effective stiffness of the modes could be a very complex
quantity with a tensor nature and could be significantly different
from the ideal value of $\sim$ 1~TPa as observed for the case of
rippled carbon nanotubes \cite{CNTrippled,theoryripledCNT,
theoryripledCNT2} and d) the capacitance to gate ($C_g$) for the
graphene membrane is likely to be mode dependent. Further
experiments with pristine unrippled graphene resonators are needed
to clarify our understanding of the microscopic origin of modes. The
presence of multiple closely spaced modes can be potentially useful
for mass spectrometry as the position of the added mass can be
extracted accurately \cite{multiModeForMassSensing}.

\section{Conclusions}
To conclude, our measurements of  $Q$, for graphene resonators, as a
function of DC gate voltage indicate that larger dissipation in the
resonator can occur around the Dirac point. Using these NEMS devices
measurement of coefficient of thermal expansion of graphene at low
temperatures is possible. Our measurements indicate that
$\alpha_{graphene}$ is negative for all temperatures between 300~K
and 30~K and larger in magnitude than the numbers predicted by
theoretical calculations \cite{expansiontheory}; this could be
critical in designing strain engineered devices using graphene
\cite{castro-neto-strainengineering,geim-graphene-strained}.
Additionally, the modal dispersion of graphene resonators is
affected by the thermal expansion of graphene reducing the
tunability. The continuum description of the mechanics of
rectangular graphene membrane is inadequate for explaining the
resonances due to the presence of non-uniform impurities and
rippling of the membrane.

\ack{The authors thank the Government of India for the financial
support. The authors would like to thank Smita Gohil and Subramanian
S. for their help during the experiments.}

\section*{References}
\bibliographystyle{unsrt}


\end{document}